# Non-linear field theory Y:
## The representation of the Yang-Mills equation as the equation of the superposition of the non-linear electromagnetic waves.

ALEXANDER G. KYRIAKOS

*Saint-Petersburg State Institute of Technology,
St.Petersburg, Russia
Present address: Athens, Greece, e-mail: agkyriak@yahoo.com*

ABSTRACT. In the present paper it is shown that the Yang-Mills equation can be represented as the equation of the superposition of the non-linear electromagnetic waves. The research of the topological characteristics of this representation allows us to discuss a number of the important questions of the quantum chromodynamics.

**1.0. Introduction**

In the paper [1] it is shown that the Dirac electron theory can be represented as the electrodynamics of the curvilinear electromagnetic wave. In the present paper we will show that this representation allows us to interpret the Yang-Mills equation as the curvilinear electromagnetic waves superposition.

*1.1. Electromagnetic forms of quantum equations*

As it is known, there are two mathematical description forms of the Dirac electron equation: spinor and bispinor.

The Dirac equations in the spinor form [2,3] are the following:

$$\begin{cases} \hat{\varepsilon}\varphi + c\hat{\vec{\sigma}}\,\hat{\vec{p}}\chi + mc^2\varphi = 0, \\ \hat{\varepsilon}\chi + c\hat{\vec{\sigma}}\,\hat{\vec{p}}\varphi - mc^2\chi = 0, \end{cases} \quad (1.1)$$

where $\hat{\vec{\sigma}}$ are Pauli matrices $\hat{\sigma}_x = \begin{pmatrix} 0 & 1 \\ 1 & 0 \end{pmatrix}, \hat{\sigma}_y = \begin{pmatrix} 0 & -i \\ i & 0 \end{pmatrix}, \hat{\sigma}_z = \begin{pmatrix} 1 & 0 \\ 0 & -1 \end{pmatrix}, \hat{\sigma}_0 = \begin{pmatrix} 1 & 0 \\ 0 & 1 \end{pmatrix}$, and $\varphi$ and $\chi$ are the so-called spinors, represented by the following matrices: $\varphi = \begin{pmatrix} \psi_1 \\ \psi_2 \end{pmatrix}, \chi = \begin{pmatrix} \psi_3 \\ \psi_4 \end{pmatrix}$.

More often the Dirac equation is described in the bispinor form, entering the function:

$$\psi = \begin{pmatrix} \varphi \\ \chi \end{pmatrix}, \quad (1.2)$$

called bispinor.

The typical Hermitian-conjugated bispinor Dirac equation forms are [2,3]:

$$\begin{aligned}[(\hat{\alpha}_o \hat{\varepsilon} + c\hat{\vec{\alpha}}\,\hat{\vec{p}}) + \hat{\beta}\,mc^2]\psi = 0 \\ \psi^+[(\hat{\alpha}_o \hat{\varepsilon} - c\hat{\vec{\alpha}}\,\hat{\vec{p}}) - \hat{\beta}\,mc^2] = 0\end{aligned} \quad (1.3)$$

where $\hat{\alpha}_0 = \begin{pmatrix} \hat{\sigma}_0 & 0 \\ 0 & \hat{\sigma}_0 \end{pmatrix}; \hat{\vec{\alpha}} = \begin{pmatrix} 0 & \hat{\vec{\sigma}} \\ \hat{\vec{\sigma}} & 0 \end{pmatrix}; \hat{\beta} \equiv \hat{\alpha}_4 = \begin{pmatrix} \hat{\sigma}_0 & 0 \\ 0 & -\hat{\sigma}_0 \end{pmatrix}$ are Dirac's matrices.

It is known for a long time [3,4] that the quantum equations can formally be represented as the Maxwell equation system. For example, according to [3] the spinor Dirac's equation system becomes the Maxwell equation system, if we put $m = 0$ and use instead of the $\varphi$ and $\chi$ 2x1-matrix wave functions, the 3x1-matrix electromagnetic field function:

$$\varphi = (\vec{E}), \quad \chi = (i\vec{H}), \quad (1.4)$$

and instead of 2x2 spinor Pauli matrices $\hat{\sigma}$ we will use the photon 3x3 spin matrices $\hat{\vec{S}}$.

*1.2. The Dirac and the Yang-Mills equations*

As it follows from the Standard Model theory [5,6] the quark family is analogue to the lepton family and the Yang-Mils equation is the generalisation of the Dirac electron equation.

The Dirac equation for the electron in the external field can be written in the form [2]:

$$\hat{\alpha}_\mu (\hat{p}_\mu + p^e_\mu)\psi + \hat{\beta}\,m_e c^2 \psi = 0 \quad (1.5)$$

where $\mu = 0,1,2,3$, $\hat{p}_\mu = \{\hat{\varepsilon}, c\hat{\vec{p}}\}$, where $\hat{\varepsilon} = i\hbar \frac{\partial}{\partial t}$, $\hat{\vec{p}} = -i\hbar \vec{\nabla}$ are the operators of energy and momentum, respectively; $\hat{p}^e_\mu = \{\hat{\varepsilon}^e_{ex}, c\hat{\vec{p}}^e_{ex}\} = j_\mu A_\mu$, where $\varepsilon^e_{ex} = e\varphi$, $\vec{p}^e_{ex} = \frac{e}{c}\vec{A}$ are the electron energy and momentum in the external electromagnetic field; respectively; $(\varphi, \vec{A})$ is 4-potential of the external field; $c$ is the light velocity, $-e$, $m$ are the electrical charge and mass of the electron correspondingly.

In Quantum Chromodynamics, which is described by Yang-Mills equation, we have quarks instead of electrons, and gluons instead of photons, between which there are the strong interactions instead of the electromagnetic interactions. The Yang-Mills equation for one quark may by written [5,6] similarly to (1.5):

$$\hat{\alpha}_\mu (\hat{p}_\mu + p^q_\mu)\psi_q + \hat{\beta}\,m_q c^2 \psi_q = 0, \quad (1.6)$$

where $\psi_i$ are the quark fields, $p^q_\mu \equiv icg\vec{G}_\mu$ with $\vec{G}_\mu = \frac{1}{2}\sum_{a=1}^{8} G^a_\mu \lambda_a$ is the potential of the gluon field, $\lambda_a, g, m_q$ are the Gell-Mann matrices, strong charge and quark mass, respectively.

*1.3. "One quark" theory of hadrons*

Formally we can say [5,6], that hadron is described by two or three Dirac electron equations of (1.1) type. Thus, conditionally we can name the Dirac electron equation as the "one quark" equation.

But here we need to take into account that the Dirac equation (1.1) is not the free electron equation. On the other hand, the equation (1.2) is indeed the equation of the "free" quark. The external field terms are used in the QED for the description of the interaction between the electron and other particles. The similar terms in the Yang-Mills equation are the internal field, describing the quark-quark interaction of the same hadron.

Further our strategy will be, as far as it is possible, to lead the analogy to the electrodynamics.

**2.0. The electromagnetic representation of "one quark" equation**

*2.1. The electron-positron pair production process*

The electron-positron pair production process can be considered as the transformation (disintegration) of the massless quantum of an electromagnetic wave $\gamma$ into two massive particles (electron-positron) $e^-, e^+$:

$$\gamma \to e^+ + e^-, \qquad (2.1)$$

Let us try to write this process in the electromagnetic form.

*2.2. The electromagnetic wave equation in the quantum form*

Let us consider the plane electromagnetic wave moving, for example, on $y$ - axis in the complex form. In the general case this electromagnetic wave has two polarizations and contains the following field vectors:

$$E_x, E_z, H_x, H_z \quad (E_y = H_y = 0), \qquad (2.2)$$

The electromagnetic wave equation has the following view [7]:

$$\left( \frac{\partial^2}{\partial t^2} - c^2 \vec{\nabla}^2 \right) \vec{F} = 0, \qquad (2.3)$$

where $\vec{F}$ is any of the electromagnetic wave fields, particularly, the fields (2.2). In other words this equation represents four equations: one for each wave function of the electromagnetic field. In the case of the wave, moving along the $y$ - axis, we can write this equation in the following form:

$$\left( \hat{\varepsilon}^2 - c^2 \hat{\vec{p}}^2 \right) \vec{F}(y) = 0, \qquad (2.4)$$

The equation (2.4) can also be represented in the form of the Klein-Gordon equation without mass:

$$\left[\left(\hat{\alpha}_o\hat{\varepsilon}\right)^2 - c^2\left(\hat{\vec{\alpha}}\ \hat{\vec{p}}\right)^2\right]\psi = 0, \qquad (2.5)$$

where $\psi$ is some matrix, which consists of the four components of $\vec{F}(y)$.

In fact, taking into account that $\left(\hat{\alpha}_o\hat{\varepsilon}\right)^2 = \hat{\varepsilon}^2$, $\left(\hat{\vec{\alpha}}\ \hat{\vec{p}}\right)^2 = \hat{\vec{p}}^2$, we see that the equations (2.4) and (2.5) are equivalent. Factorising (2.5) and multiplying it from the left on the Hermitian-conjugate function $\psi^+$ we get:

$$\psi^+\left(\hat{\alpha}_o\hat{\varepsilon} - c\hat{\vec{\alpha}}\ \hat{\vec{p}}\right)\left(\hat{\alpha}_o\hat{\varepsilon} + c\hat{\vec{\alpha}}\ \hat{\vec{p}}\right)\psi = 0, \qquad (2.6)$$

The equation (2.6) may be disintegrated on two Dirac equations without mass:

$$\begin{aligned}\psi^+\left(\hat{\alpha}_o\hat{\varepsilon} - c\hat{\vec{\alpha}}\ \hat{\vec{p}}\right) &= 0 \\ \left(\hat{\alpha}_o\hat{\varepsilon} + c\hat{\vec{\alpha}}\ \hat{\vec{p}}\right)\psi &= 0\end{aligned}, \qquad (2.7)$$

It is not difficult to show that only in the case when we choose the Dirac bispinors in the following way:

$$\psi = \begin{pmatrix} E_x \\ E_z \\ iH_x \\ iH_z \end{pmatrix},\ \psi^+ = \begin{pmatrix} E_x & E_z & -iH_x & -iH_z \end{pmatrix}, \qquad (2.8)$$

the equations (2.7) are the right Maxwell equations of the electromagnetic waves: retarded and advanced. (The same results can be obtained for waves of any other direction by the cyclic transposition of the indexes and by the canonical transformation of matrices and wave functions [8]).

Then the question arises: which transformation can turn the equations of the massless electromagnetic wave (2.7) into the Dirac equations (1.3) of the electron and positron with mass terms?

We will show that it can be made, at least, in two ways: either by using the curvilinear metrics, or by using differential geometry.

*2.3. The introduction of the mass term by using of the curvilinear metrics*

The generalization of the Dirac equation on the curvilinear (Riemann) geometry is connected with the parallel transport of the spinor in the curvilinear space [9,10]. We will use the most important results of this theory below.

For the generalization of the Dirac equation on the Riemann geometry it is necessary to replace the usual derivative $\partial_\mu \equiv \partial/\partial x_\mu$ (where $x_\mu$ are the co-ordinates in the 4-space) with the covariant derivative:

$$D_\mu = \partial_\mu + \Gamma_\mu, \qquad (2.9)$$

where $\mu = 0, 1, 2, 3$ are the summing indices and $\Gamma_\mu$ is the analogue of Christoffel's symbols in the case of the spinor theory, called Ricci connection coefficients. In the theory it shown that $\hat{\alpha}_\mu \Gamma_\mu = \hat{\alpha}_i p_i + i\hat{\alpha}_0 p_0$, where $p_i$ and $p_0$ are the real values.

When a spinor moves along the straight line, all the $\Gamma_\mu = 0$ and we have a usual derivative. But if a spinor moves along the curvilinear trajectory, not all the $\Gamma_\mu$ are equal to zero and a supplementary term appears. Typically, the last one is not the derivative, but it is equal to the product of the spinor itself with some coefficient $\Gamma_\mu$. It is not difficult to show that the supplementary term contains a mass. Since, according to the general theory [9], the increment in spinor $\Gamma_\mu$ has the form and the dimension of the energy-momentum 4-vector, it is logical to identify $\Gamma_\mu$ with 4-vector of energy-momentum of the photon electromagnetic field:

$$\Gamma_\mu = \{\varepsilon_p, c\vec{p}_p\}, \qquad (2.10)$$

where $\varepsilon_p$ and $p_p$ is the photon energy and momentum (not the operators). Then the equations (2.7) in the curvilinear space will have the form:

$$\begin{aligned}[(\hat{\alpha}_o\hat{\varepsilon} + c\hat{\vec{\alpha}}\ \hat{\vec{p}}) + (\hat{\alpha}_o\varepsilon_p + c\hat{\vec{\alpha}}\ \vec{p}_p)]\psi = 0 \\ \psi^+[(\hat{\alpha}_o\hat{\varepsilon} - c\hat{\vec{\alpha}}\ \hat{\vec{p}}) - (\hat{\alpha}_o\varepsilon_p - c\hat{\vec{\alpha}}\ \vec{p}_p)] = 0\end{aligned}, \qquad (2.11)$$

According to the energy conservation law we can formally write:

$$\hat{\alpha}_o\varepsilon_p \pm c\hat{\vec{\alpha}}\ \vec{p}_p = \mp\hat{\beta}\,m\,c^2, \qquad (2.12)$$

Substituting (2.12) in (2.11) we will arrive at the usual kind of two Dirac Hermitian-conjugate equations with the mass (1.3)

Now we consider the electromagnetic representation of the Dirac equation mass term.

*2.4. Electrodynamics form of the "one quark" equation with mass*

Using (2.8) from (1.3) we obtain:

$$\begin{cases}\dfrac{1}{c}\dfrac{\partial E_x}{\partial t} - \dfrac{\partial H_z}{\partial y} = -\vec{j}_x^e, \\ \dfrac{1}{c}\dfrac{\partial H_z}{\partial t} - \dfrac{\partial E_x}{\partial y} = \vec{j}_z^m, \\ \dfrac{1}{c}\dfrac{\partial E_z}{\partial t} + \dfrac{\partial H_x}{\partial y} = -\vec{j}_z^e, \\ \dfrac{1}{c}\dfrac{\partial H_x}{\partial t} + \dfrac{\partial E_z}{\partial y} = \vec{j}_x^m,\end{cases} (2.13), \quad \begin{cases}\dfrac{1}{c}\dfrac{\partial E_x}{\partial t} - \dfrac{\partial H_z}{\partial y} = \vec{j}_x^e, \\ \dfrac{1}{c}\dfrac{\partial H_z}{\partial t} - \dfrac{\partial E_x}{\partial y} = -\vec{j}_z^m, \\ \dfrac{1}{c}\dfrac{\partial E_z}{\partial t} + \dfrac{\partial H_x}{\partial y} = \vec{j}_z^e, \\ \dfrac{1}{c}\dfrac{\partial H_x}{\partial t} + \dfrac{\partial E_z}{\partial y} = -\vec{j}_x^m,\end{cases} (2.14)$$

where

$$\vec{j}^e = i\frac{\omega}{4\pi}\vec{E} = i\frac{c}{4\pi}\frac{1}{r_C}\vec{E}$$
$$\vec{j}^m = i\frac{\omega}{4\pi}\vec{H} = i\frac{c}{4\pi}\frac{1}{r_C}\vec{H}$$
(2.15)

are the imaginary currents, in which $\omega = \frac{mc^2}{\hbar}$, and $r_C = \frac{\hbar}{mc}$ is the Compton length wave of the electron. Thus, the equations (2.13) and (2.14) are Maxwell equations with imaginary currents, which differ by the directions. (As it is known the existence of the magnetic current $\vec{j}^m$ doesn't contradict to the quantum theory; see the Dirac theory of the magnetic monopole [11]).

Now we will consider the physical origin of the appearance of the current-mass term from the differential geometry point of view.

*2.5. Electromagnetic origin of the current-mass term appearance*

Let us show that the current-mass term (2.23) is the additional Maxwell displacement current that appears when the initial electromagnetic wave changes its trajectory from the linear to the curvilinear. Noting that the Pauli matrices are the generators of the rotation transformations in the 2D-space, we can
suppose that this curvilinear trajectory is plane.

Let the plane-polarized wave, which has the field vectors $(E_x, H_z)$, be twirled with some radius $r_K$ in the plane $(X', O', Y')$ of a fixed co-ordinate system $(X', Y', Z', O')$ so that $E_x$ is parallel to the plane $(X', O', Y')$ and $H_z$ is perpendicular to it.

According to Maxwell [7] the displacement current is defined by the equation:

$$j_{dis} = \frac{1}{4\pi}\frac{\partial \vec{E}}{\partial t},$$
(2.16)

The above electrical field vector $\vec{E}$, which moves along the curvilinear trajectory (let it have direction from the centre), can be written in the form:

$$\vec{E} = -E \cdot \vec{n},$$
(2.17)

where $E = |\vec{E}|$, and $\vec{n}$ is the normal unit-vector of the curve (having direction to the centre). The derivative of $\vec{E}$ can be represented as:

$$\frac{\partial \vec{E}}{\partial t} = -\frac{\partial E}{\partial t}\vec{n} - E\frac{\partial \vec{n}}{\partial t},$$
(2.18)

Here the first term has the same direction as $\vec{E}$. The existence of the second term shows that the additional displacement current appears at the twirling of the wave. It is not difficult to show that it has direction, tangential to the ring:

$$\frac{\partial \vec{n}}{\partial t} = -\upsilon_p K \vec{\tau},  \qquad (2.19)$$

where $\vec{\tau}$ is the tangential unit-vector, $\upsilon_p \equiv c$ is the electromagnetic wave velocity, $K = \dfrac{1}{r_K}$ is the curvature of the trajectory and $r_K$ is the curvature radius. Thus, the displacement current of the plane wave, moving along the ring, can be written in the form:

$$\vec{j}_{dis} = -\frac{1}{4\pi}\frac{\partial E}{\partial t}\vec{n} + \frac{1}{4\pi}\omega_K E \cdot \vec{\tau}, \qquad (2.20)$$

where $\omega_K = \dfrac{\upsilon_p}{r_K} \equiv cK$ we name the curvature angular velocity, $\vec{j}_n = \dfrac{1}{4\pi}\dfrac{\partial E}{\partial t}\vec{n}$ and $\vec{j}_\tau = \dfrac{\omega_K}{4\pi}E\cdot\vec{\tau}$ are the normal and tangent components of the current of the twirled electromagnetic wave, correspondingly. Here the current $\vec{j}_\tau = \dfrac{\omega_K}{4\pi}E\cdot\vec{\tau}$ corresponds to the electrical current (2.15), if we take into account the fact that the imaginary unit correspondents to the rotation on $\pi/2$ and $r_K = r_C$. (Note that it can be shown that when the electromagnetic wave has the circular polarization the magnetic current (2.15) also appears, but integrally this current is equal to zero).

## 3.0. Electromagnetic representation of Yang-Mills equation

Obviously, to obtain the Yang-Mills equation we must sum three "one quark" equations without mass and "turn on" the self-interactions among quarks.

*3.1. Electromagnetic forms of "three quark" equations*

As the Pauli matrices are [12] the generators of the 2D rotation, for the "three quark" electromagnetic representation we must use the generators of the 3D rotation, which are the known photon spin 3x3-matrices $\hat{\vec{S}}$ of the O(3) group [3,12]:

$$\hat{S}_1 = \begin{pmatrix} 0 & 0 & 0 \\ 0 & 0 & -i \\ 0 & i & 0 \end{pmatrix}, \hat{S}_2 = \begin{pmatrix} 0 & 0 & i \\ 0 & 0 & 0 \\ i & 0 & 0 \end{pmatrix}, \hat{S}_3 = \begin{pmatrix} 0 & -i & 0 \\ i & 0 & 0 \\ 0 & 0 & 0 \end{pmatrix}, \qquad (3.1)$$

As the "three quark" equations for the particle and antiparticle we will use the Dirac equations (1.1) in the following form:

$$\begin{aligned}\left[\left({}^6\hat{\alpha}_o\hat{\varepsilon} - c{}^6\hat{\vec{\alpha}}\;\hat{\vec{p}}\right) - {}^6\hat{\beta}\,mc^2\right]\psi &= 0 \\ \psi^+\left[\left({}^6\hat{\alpha}_o\hat{\varepsilon} + c{}^6\hat{\vec{\alpha}}\;\hat{\vec{p}}\right) + {}^6\hat{\beta}\,mc^2\right] &= 0\end{aligned}, \qquad (3.2)$$

where the left upper index "6" means that these matrices are the 6x6-matrices of the following type:

$$^6\hat{\vec{\alpha}} = \begin{pmatrix} \hat{0} & \hat{\vec{S}} \\ \hat{\vec{S}} & \hat{0} \end{pmatrix}, \quad ^6\hat{\alpha}_0 = \begin{pmatrix} \hat{S}_0 & \hat{0} \\ \hat{0} & \hat{S}_0 \end{pmatrix}, \quad ^6\hat{\alpha}_4 \equiv {}^6\hat{\beta} = \begin{pmatrix} \hat{S}_0 & \hat{0} \\ \hat{0} & -\hat{S}_0 \end{pmatrix}, \quad (3.3)$$

Here $\hat{S}_0 = \hat{1}$ and wave function $^6\psi = \begin{pmatrix} \vec{E} \\ i\vec{H} \end{pmatrix}$ is the 6x1 matrix.

As it is not difficult to test if the above matrices give the right electromagnetic expressions of the bilinear form of the theory: of the energy $^6\psi^+ \, ^6\hat{\alpha}_0 \, ^6\psi = \vec{E}^2 + \vec{H}^2 = 8\pi U$, of the Poynting vector $\vec{S}_P = \frac{1}{8\pi} {}^6\psi^+ \, ^6\vec{\alpha} \, ^6\psi$, and of the 1$^{st}$ scalar of the electromagnetic field: $^6\psi^+ \, ^6\hat{\alpha}_4 \, ^6\psi = 2(\vec{E}^2 - \vec{H}^2) = 4\pi \, F_{\mu\nu} F^{\mu\nu}$.

*3.2. "Three-quarks" equation without interaction*

From the above follows that the proton equation can be represented by three "one quark" equations, i.e. three electron equations or three pairs of the scalar Maxwell equations (one pair for each co-ordinate). Obviously, there is a possibility of two directions of rotations of each quark (the left and the right quarks). Therefore, the 6+6 scalar equations for proton description must exist as well as the 6+6 equations for the antiproton description.

Let us find these equations without interaction, putting the interaction (mass) terms equal to zero. Using (3.3) from the equations (3.2) we obtain the Maxwell equation without current:

$$\begin{cases} \frac{1}{c}\frac{\partial E_x}{\partial t} - \frac{\partial H_z}{\partial y} = 0, & a \\ \frac{1}{c}\frac{\partial H_z}{\partial t} - \frac{\partial E_x}{\partial y} = 0, & a' \\ \frac{1}{c}\frac{\partial E_y}{\partial t} - \frac{\partial H_x}{\partial z} = 0, & b \\ \frac{1}{c}\frac{\partial H_x}{\partial t} - \frac{\partial E_y}{\partial z} = 0, & b' \\ \frac{1}{c}\frac{\partial E_z}{\partial t} - \frac{\partial H_y}{\partial x} = 0, & c \\ \frac{1}{c}\frac{\partial H_y}{\partial t} - \frac{\partial E_z}{\partial x} = 0, & c' \end{cases} (3.4) \quad \begin{cases} \frac{1}{c}\frac{\partial E_z}{\partial t} + \frac{\partial H_x}{\partial y} = 0, & a \\ \frac{1}{c}\frac{\partial H_x}{\partial t} + \frac{\partial E_z}{\partial y} = 0, & a' \\ \frac{1}{c}\frac{\partial E_x}{\partial t} + \frac{\partial H_y}{\partial z} = 0, & b \\ \frac{1}{c}\frac{\partial H_y}{\partial t} + \frac{\partial E_x}{\partial z} = 0, & b' \\ \frac{1}{c}\frac{\partial E_y}{\partial t} + \frac{\partial H_z}{\partial x} = 0, & c \\ \frac{1}{c}\frac{\partial H_z}{\partial t} + \frac{\partial E_y}{\partial x} = 0, & c' \end{cases} (3.5)$$

As it is not difficult to see that each pair of the equations *a,b,c* describes a separate ring; the fields vectors of equations (3.4) are rolled up in the plains *XOZ, ZOY, YOX*, and similarly the fields vectors of the equations (3.5) are rolled in the plains *XOY, YOZ, ZOX*.

*3.3. The interaction appearance*

The modern particle theory is also known as the gauge field theory because the interactions between the particles are introduced in the field equation via the gauge transformations. It is known [12,13] that this procedure is mathematically equivalent to the field vector transformations in the curvilinear space, which lead to the covariant derivative appearance.

It is not difficult to show [12], that the electromagnetic field appears naturally as a consequence of the requirement of the Lagrangian invariance relatively to the gauge transformations of the local rotations in the internal space of the complex field $\psi$, when the Lagrangian has the symmetry O(2) or U(1). Mathematically this is expressed through the replacement of the simple derivatives with the covariant derivatives.

The generalization of this result on a case of 3D-space is the Yang - Mills field. The elementary generalization of this symmetry is the non-abelian group SU (2); i.e. the question is about the theory of the non-abelian gauge fields.

We shall consider the rotation of some field vector $\vec{F}$ in three-dimensional space around some axis on an infinitesimal corner. Here the size $|\vec{\varphi}|$ is a rotation corner, and the vector $\vec{\varphi}/|\vec{\varphi}|$ sets the direction of the axis of rotation. The transition from the initial position of a vector to the final position will be defined by the transformation:

$$\vec{F} \to \vec{F}' = \vec{F} - \vec{\varphi} \times \vec{F}, \qquad (3.6)$$

The problem is, that it is necessary to create independent rotations in various points of space. In order to construct correctly the covariant derivative of a field, we should make parallel transport of the vectors into the space, instead of on a flat curve, as in the above case of spinorial theory. The corresponding analysis [12] allows us to receive an expression similar to the expression about the spinor transport on a flat curve.

It can be shown [12] that the expression

$$\frac{D\psi}{dx^\mu} = D_\mu \psi = \left( \partial_\mu - ig M^a A^a_\mu \right)\psi, \qquad (3.7)$$

defines a covariant derivative of the field $\psi$, which is transformed according to some representation of a group (here the matrixes $M^a$ are the generators of the rotation). It is not difficult to make sure, that this expression gives the same covariant derivative, as found earlier, and can give the mass terms.
In the following section we will consider the electromagnetic discription of the mass term appearance.

*3.4.The electromagnetic description of the interaction appearance*

The spinorial theory shows that the appearance of the internal interaction terms is bounded with the three vectors $\vec{E}, \vec{H}, \vec{S}_P$, moving along the curvilinear trajectory. These vectors represent the moving trihedral of the Frenet-Serret [14]. In the general case, when the electromagnetic wave field vectors of three-quark particles move along the space curvilinear trajectories, not only the additional term, defined by the curvature, appears, but also the terms that are defined by the torsion of the trajectory.

Actually, in this case we have:

$$\frac{\partial \vec{E}}{\partial t} = -\frac{\partial E}{\partial t}\vec{n} - E\frac{\partial \vec{n}}{\partial t}$$

$$\frac{\partial \vec{H}}{\partial t} = \frac{\partial \vec{H}}{\partial t}\vec{b} + H\frac{\partial \vec{b}}{\partial t}$$

(3.8)

where $\vec{b}$ is the binormal vector. According to the Frenet-Serret formulas we have:

$$\frac{\partial \vec{n}}{\partial t} = -\upsilon_p K\vec{\tau} + \upsilon_p T\vec{b}$$

$$\frac{\partial \vec{b}}{\partial t} = -\upsilon_p T\vec{n}$$

(3.9)

where $T = \frac{1}{r_T}$ is the torsion of the trajectory and $r_T$ is the torsion radius. Thus, the displacement currents can be written in the form:

$$\vec{j}^e = -\frac{1}{4\pi}\frac{\partial E}{\partial t}\vec{n} + \frac{1}{4\pi}\omega_K E \cdot \vec{\tau} - \frac{1}{4\pi}\omega_T E \cdot \vec{b}$$

$$\vec{j}^m = \frac{1}{4\pi}\frac{\partial H}{\partial t}\vec{b} - \frac{1}{4\pi}\omega_T H \cdot \vec{n}$$

(3.10)

where $\omega_T = \frac{\upsilon_p}{r_T} \equiv cT$ we name the torsion angular velocity.

Thus, we can obtain the following electromagnetic representation of the Yang-Mills equations:

$$\begin{cases} \frac{1}{c}\frac{\partial E_x}{\partial t} - \frac{\partial H_z}{\partial y} = j_1^e, & a \\ \frac{1}{c}\frac{\partial H_z}{\partial t} - \frac{\partial E_x}{\partial y} = j_1^m, & a' \\ \frac{1}{c}\frac{\partial E_y}{\partial t} - \frac{\partial H_x}{\partial z} = j_2^e, & b \\ \frac{1}{c}\frac{\partial H_x}{\partial t} - \frac{\partial E_y}{\partial z} = j_2^m, & b' \\ \frac{1}{c}\frac{\partial E_z}{\partial t} - \frac{\partial H_y}{\partial x} = j_3^e, & c \\ \frac{1}{c}\frac{\partial H_y}{\partial t} - \frac{\partial E_z}{\partial x} = j_3^m, & c' \end{cases} (3.11) \begin{cases} \frac{1}{c}\frac{\partial E_z}{\partial t} + \frac{\partial H_x}{\partial y} = j_1^e, & a \\ \frac{1}{c}\frac{\partial H_x}{\partial t} + \frac{\partial E_z}{\partial y} = j_1^m, & a' \\ \frac{1}{c}\frac{\partial E_x}{\partial t} + \frac{\partial H_y}{\partial z} = j_2^e, & b \\ \frac{1}{c}\frac{\partial H_y}{\partial t} + \frac{\partial E_x}{\partial z} = j_2^m, & b' \\ \frac{1}{c}\frac{\partial E_y}{\partial t} + \frac{\partial H_z}{\partial x} = j_3^e, & c \\ \frac{1}{c}\frac{\partial H_z}{\partial t} + \frac{\partial E_y}{\partial x} = j_3^m, & c' \end{cases} (3.12)$$

where $j_k$ ($k = 1,2,3$) are the currents of each quark:

$$\vec{j}_k^e = i\frac{\omega_k}{4\pi}\vec{E} \text{ and } \vec{j}_k^m = i\frac{\omega_k}{4\pi}\vec{H}, \qquad (3.13)$$

Thus, in the framework of the electromagnetic representation a proton is topologically the superposition of the three rings. Therefore, we can suppose that in this representation it has the scheme of the trefoil knot (fig. 1):

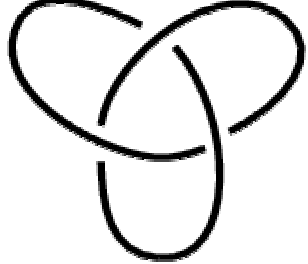

Fig.1

(see also fig. from [15], where the animation shows a series of gears arranged along a Möbius strip trefoil knot as the electric and magnetic field vectors motion).

**Discussion**

The above electromagnetic representation of the Yang-Mills equations allows us to discuss some particularities of the QCD:

1. *The fractional charge of the quarks*: according to the above results the electric field trajectory of the quarks, not only has a curvature, but also torsion; hence, the tangential current, generated by the electrical field vector transport, alternates along the space trajectory. Consequently, the electric charge of one knot, as an integral from this current, will be less, than the electron charge. But the total charge from all knots can be equal to the charge of the electron.

2. *Quarks confinement*: since quarks are three connected knots, they cannot exist in a free state.

3. *The masses of the quarks* are defined by the rotation frequencies of each knot. It is possible to show, that the figure, formed from three knots, forms a steady construction only at the certain circular frequencies.

4. *Non-linearity of the Yang-Mills equation*: obviously, the Yang-Mills equation as the non-linear electromagnetic waves superposition is the non-linear equation.

5. The g*luons and photons analogy*: according to the topological model of the proton, the gluons are the virtual photons, by which the knots interact between themselves.

A question arises of whether the quantum and the electromagnetic forms belong to the same theory or it is a mere coincidence.